\documentclass[aps,prx,twocolumn,floatfix,longbibliography,superscriptaddress]{revtex4-2}

\usepackage{graphicx}
\usepackage{amsmath}
\usepackage{amssymb}
\usepackage{braket}
\usepackage{bbold}
\usepackage{hyperref}
\usepackage{subfigure}
\usepackage[usenames,dvipsnames]{color}
\usepackage{graphicx}
\usepackage{subfigure}
\usepackage[mathscr]{euscript}
\hypersetup{colorlinks,allcolors=black}

\begin{document}
\title{Topological Phase Transition under Infinite Randomness}

\author{Saikat Mondal}
\email{msaikat@iitk.ac.in}
\affiliation{Department of Physics, Indian Institute of Technology Kanpur, Kalyanpur, Uttar Pradesh 208016, India}

\author{Adhip Agarwala}
\email{adhip@iitk.ac.in}
\affiliation{Department of Physics, Indian Institute of Technology Kanpur, Kalyanpur, Uttar Pradesh 208016, India}

\begin{abstract}
In clean and weakly disordered systems, topological and trivial phases having a finite bulk energy gap can transit to each other via a quantum critical point. In presence of strong disorder, both the nature of the phases and the associated criticality can fundamentally change. Here we investigate topological properties of a strongly disordered fermionic chain where the bond couplings are drawn from normal probability distributions which are defined by characteristic standard deviations. Using numerical strong disorder renormalization group methods along with analytical techniques, we show that the 
competition between fluctuation scales renders both the trivial and topological phases gapless with Griffiths' like rare regions. 
Moreover, the transition between these phases is solely governed by the fluctuation scales, rather than the means, rendering the critical behavior to be determined by an infinite randomness fixed point with an irrational central charge.  Our work points to a host of novel topological phases and atypical topological phase transitions which can be realized in systems under strong disorder. 
\end{abstract}

\maketitle

{\it Introduction:} Nature of topological phases with and without disorder has been a subject of intense scrutiny given both their technological implications and the effective theories they realize~\cite{hasan_rmp_2010,lukasz11,qi_rmp_2011,prodan_jpamt_2011,chen13,senthil15,chiu_rmp_2016}. Even in one-dimensional topological phases, disorder has been shown to illustrate non-trivial criticality and rich phase diagrams~\cite{altland_prl_2014,prodan_prl_2014,prodan_prb_2014,liu_physletta_2018,perez_prb_2019,hsu_prb_2020,shi_prr_2021,cinnirella_prb_2024,kar_prb_2024,sircar_arxiv_2024,liu_natcom_2025,lewenstein_mlst_2025}. Given the experimental realizations in host of cold atomic, electronic and mechanical platforms, some of these studies are now amenable to direct experimental verification~\cite{meier_science_2018,longhi_optlett_2020,thatcher_physcr_2022,splitthoff_prr_2024}.  However, in most of these systems, at large disorder one eventually realizes a featureless Anderson insulator where the disorder scale overwhelms all other energy scales. In spin systems, strong disorder is known to generate Griffiths' phases and Infinite Randomness Fixed Point (IRFP)~\cite{fisher_prl_1992,fisher_prb_1994,fisher_prb_1995,young_prb_1996,rafael_prl_2004,nicolas_prb_2005,melin_prb_2005,hoyos_prb_2007,refael_jpmt_2009,fagotti_prb_2011,filho_jsmte_2012,igloi_epjb_2018,tsai_epjb_2020,alcaraz_prb_2021,huse_arxiv_2023,alcaraz_prb_2023,kovacs_prb_2024}. In this context, a particularly illuminating method -- Dasgupta-Ma strong-disorder renormalization group (SDRG) method was developed~\cite{ma_prl_1979,dasgupta_prb_1980,igloi_physrep_2005,rafael_crphysique_2013}. In such highly disordered systems, often termed infinitely random systems, {\it  competing} fluctuation (or disorder) scales can lead to highly non-trivial physics~\cite{fisher_prl_1992,young_prb_1996,rafael_prl_2004}. Such physics has been little explored in context of fermionic systems~\cite{javan_prb_2014,javan_prb_2017}.  In this work we fill this lacuna by studying the paradigmatic model of a one-dimensional topological system, the Su–Schrieffer–Heeger (SSH) chain~\cite{ssh_prl_1979}, and study its behavior in context of infinitely strong disorder.

\begin{figure}
    \centering
\includegraphics[width=1.0\columnwidth]{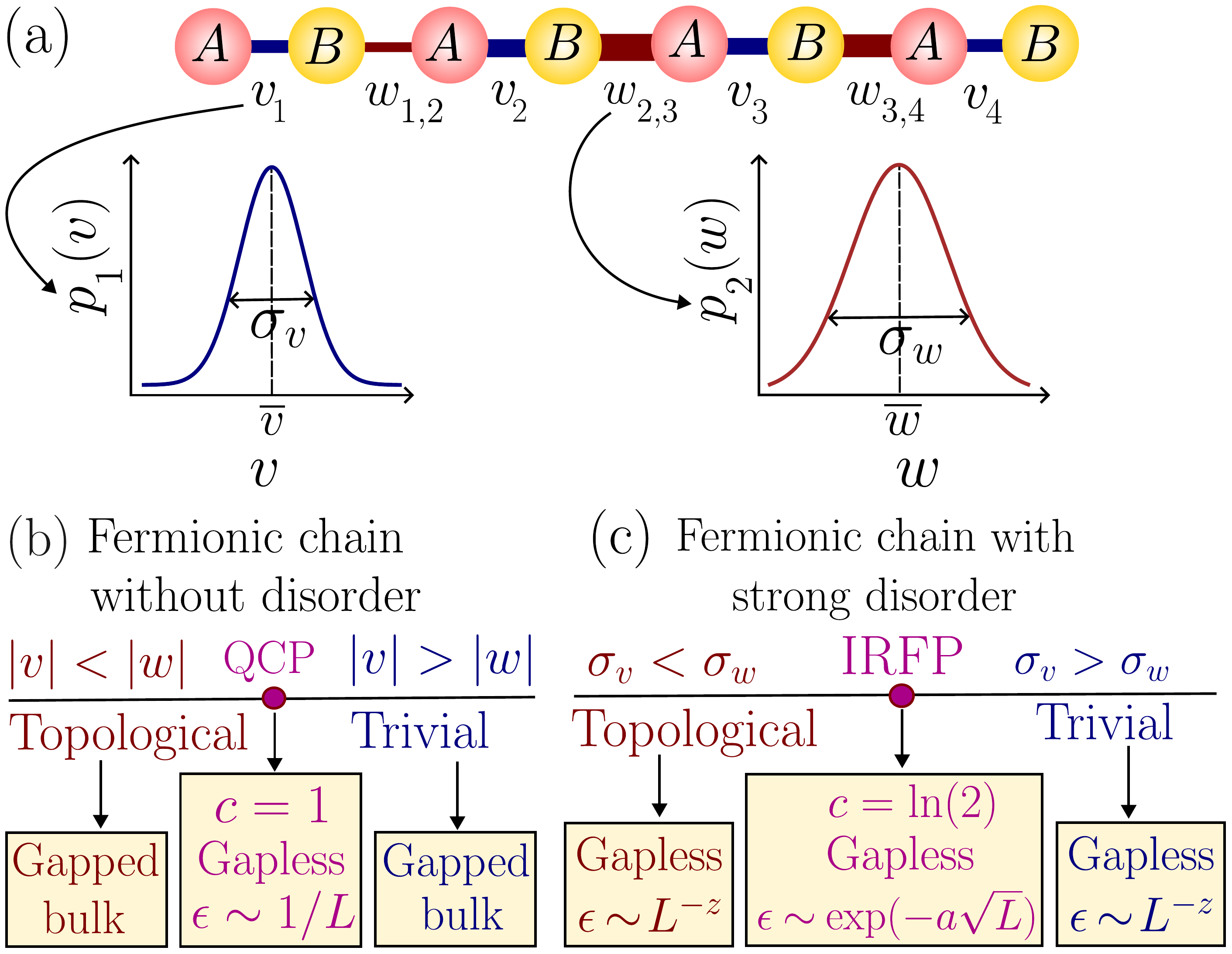}
\caption{{\textbf{Topological phase transitions without and with strong disorder:}} (a) Disordered SSH chain having $A$ and $B$ sites where intra-cell ($v$) and inter-cell ($w$) hopping parameters are chosen from normal probability distributions $p_{1}(v)$ and $p_{2}(w)$ with mean ${\overline{v}}$, ${\overline{w}}$ and standard deviation $\sigma_{v}$, $\sigma_{w}$. (b) Topological ($|v|<|w|$) and trivial ($|v|>|w|$) phases of clean SSH chain where bulk energy-gap remains finite. At quantum critical point (QCP) $|w|=|v|$, bulk energy-gap closes and central charge $c=1$. For finite system-size $L$, energy-gap at QCP is $\epsilon \sim 1/L$. (c) Topological ($\sigma_{v}<\sigma_{w}$) and trivial ($\sigma_{v}>\sigma_{w}$) phases of a strongly disordered SSH chain. These two phases are gapless ($\epsilon \sim L^{-z}$ where $z>0$) and separated by an infinite randomness fixed point (IRFP) where effective central charge is $c=\ln(2)$ and energy-gap shows activated dynamical scaling $\epsilon \sim \exp(-a \sqrt{L})$ for finite size $L$ (where $a>0$ is a constant).}
\label{fig_clean_disordered}
\end{figure}

In this work, we ask the question: can strong disorder lead to topological phase transition via a IRFP? To answer this question, we study the SSH chain with random intra-cell ($v_{n}$) and inter-cell ($w_{n,n+1}$) hopping parameters. In particular, we choose these hopping parameters from two normal probability distributions $p_{1}(v)$ and $p_{2}(w)$ that have means $\overline{v}$, $\overline{w}$ and standard deviations $\sigma_{v}$, $\sigma_{w}$ (see Fig.~\ref{fig_clean_disordered}(a)). Here, the situation with $\sigma_{v}=0$, $\sigma_{w}=0$ corresponds to clean SSH chain with a finite bulk energy gap in topological ($|v|<|w|$) and trivial ($|v|>|w|$) phases, while at the quantum critical point (QCP) $|v|=|w|$, energy gap falls as $\epsilon \sim 1/L$ for finite system size $L$ and central charge is $c=1$ (see Fig.~\ref{fig_clean_disordered}(b)). In strongly disordered chain (with finite $\sigma_{w}$), we show that transition from a topological to a trivial phase can occur by tuning the ratio $\lambda=\sigma_{v}/\sigma_{w}$, regardless of the value of ${\overline{w}}/{\overline{v}}$, where the chain is topological when $\sigma_{v}<\sigma_{w}$ and trivial when $\sigma_{v}>\sigma_{w}$. Interestingly, we find that both topological and trivial phases in strongly disordered chain are gapless, where bulk energy gap for a finite $L$ falls as $\epsilon \sim L^{-z}$ (where $z>0$). The QCP at $\sigma_{w}=\sigma_{v}$, where topological phase transition occurs, turns out to be an IRFP where energy gap falls as $\epsilon \sim \exp(-a \sqrt{L})$ where $a>0$ and effective central charge is $c=\ln(2)$ (see Fig.~\ref{fig_clean_disordered}(c)). Using numerical SDRG and analytical methods, we confirm strong disorder-driven topological phase transition occurring at an IRFP.

{\it Model:} We consider a SSH chain~\cite{ssh_prl_1979,ssh_prb_1980} represented by the Hamiltonian
\begin{align}\label{eq_disordered_ssh}
H &= \sum_{n=1}^{L} v_{n} \left( c_{n,A}^{\dagger} c_{n,B} + c_{n,B}^{\dagger} c_{n,A} \right) \nonumber \\
&+ \sum_{n=1}^{L-1} w_{n,n+1} \left( c_{n,B}^{\dagger} c_{n+1,A} + c_{n+1,A}^{\dagger} c_{n,B} \right),
\end{align}
where there are $L$ unit cells in the chain and there are $A,B$ sites in each unit cell. $c^{\dagger}_{n,A}$ ($c_{n,B}$) is fermionic creation (annihilation) operator of site $A$ ($B$) in $n$-th unit cell. $v_{n}$ is intra-cell hopping parameter in $n$-th unit cell and $w_{n,n+1}$ is inter-cell hopping parameter between $n$-th and $(n+1)$-th unit cells, where $v_{n}$ and $w_{n,n+1}$ for different $n=1,2,...,L$ are chosen from normal probability distributions with $({\rm{mean}}, {\rm{standard ~deviation}}) = ({\overline{v}}, \sigma_{v})$ and $({\overline{w}}, \sigma_{w})$ respectively. The Hamiltonian in Eq.~\eqref{eq_disordered_ssh} has time-reversal, sub-lattice, particle-hole symmetry and belongs to symmetry class BDI. We now employ SDRG methods to analyze the phases in disordered SSH chain for the parameters ${\overline{w}}/{\overline{v}}$, $\sigma_{w}$ and $\lambda= \sigma_{v}/\sigma_{w}$.

\begin{figure}
    \centering
\includegraphics[width=1.0\linewidth]{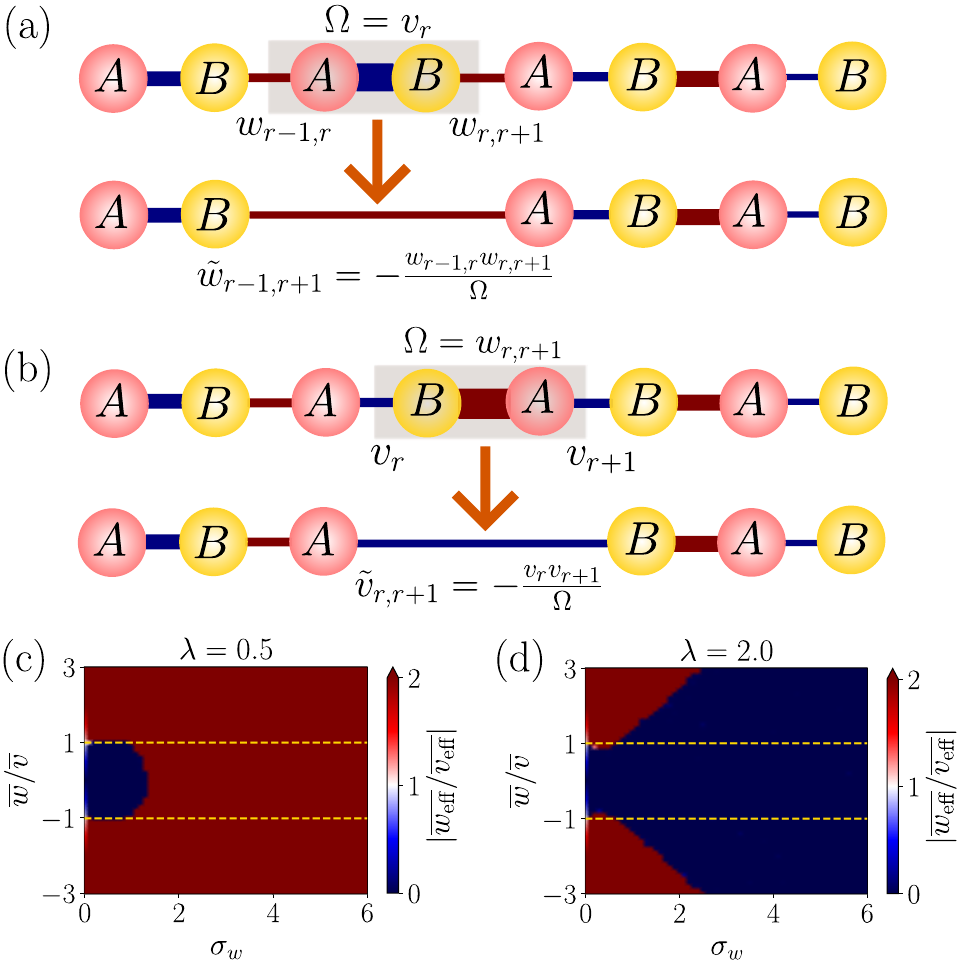}
\caption{{\textbf{Strong disorder renormalization group (SDRG) results:}} (a,b) Decimation of bonds when largest hopping strength is (a) intra-cell hopping parameter $\Omega=v_{r}$, (b) inter-cell hopping parameter $\Omega=w_{r,r+1}$. (c,d) $|\overline{w_{\rm{eff}}}/\overline{v_{\rm{eff}}}|$ calculated using SDRG methods for the number of iterations $N=15$ (see text) as a function of ${\overline{w}}/{\overline{v}}$ and $\sigma_{w}$ when (c) $\lambda=0.5$, (d) $\lambda=2.0$. For large $\sigma_{w}$, we find $|\overline{w_{\rm{eff}}}|>|\overline{v_{\rm{eff}}}|$ when $\lambda<1$, while $|\overline{w_{\rm{eff}}}|<|\overline{v_{\rm{eff}}}|$ when $\lambda>1$. In (c,d), number of realizations considered is $100$.}
\label{fig_sdrg}
\end{figure}

{\it SDRG methods:} To investigate the properties of strongly disordered chain, we resort to SDRG methods~\cite{dasgupta_prb_1980,fisher_prl_1992,javan_prb_2014}. Here, we consider periodic boundary condition where the intra-cell hopping parameters are $\{ v_{1},~v_{2},...,~v_{L-1}, ~v_{L}\}$ and the inter-cell hopping parameters are $\{ w_{1,2},~w_{2,3},...,~w_{L-1,L}, ~w_{L,0} \}$. Therefore, the numbers of intra-cell and inter-cell hopping parameters are $L$. Decimation of these hopping parameters is then carried out following the protocols given below:
\begin{enumerate}
    \item We choose the largest hopping parameter $\Omega = {\rm{max}} \{v_{n},~w_{n,n+1} \}$ among all $v_{n}$, $w_{n,n+1}$.
    \item (A) If $\Omega$ is a intra-cell hopping parameter, say $\Omega=v_{r}$, then $v_{r}$ is decimated. The inter-cell hopping parameters $w_{r-1,r}$ and $w_{r,r+1}$ associated with $r$-th unit cell are also decimated and an effective inter-cell hopping parameter
    \begin{equation}
        {\tilde{w}_{r-1,r+1}}= -\frac{w_{r-1,r} w_{r,r+1}}{\Omega}
    \end{equation}
    between $(r-1)$-th and $(r+1)$-th unit cell is included~\cite{javan_prb_2014}. Thus, the numbers of both intra-cell and inter-cell hopping parameters effectively decrease by $1$ (see Fig.~\ref{fig_sdrg}(a)).
    
    (B) If $\Omega$ is an intercell hopping parameter, say $\Omega=w_{r,r+1}$, then $w_{r,r+1}$ is decimated. The intracell hopping parameters $v_{r}$ and $v_{r+1}$ associated with $r$-th and $(r+1)$-th unit cells are also decimated and an effective hopping parameter
    \begin{equation}
        {\tilde{v}_{r,r+1}}= -\frac{v_{r} v_{r+1}}{\Omega}
    \end{equation}
    is included~\cite{javan_prb_2014}. Thus, the numbers of both intra-cell and inter-cell hopping parameters effectively decrease by $1$ (see Fig.~\ref{fig_sdrg}(b)).
    \item We continue decimation of the remaining hopping parameters as mentioned in point 2 for $L/2$ steps until the numbers of both intra-cell and inter-cell hopping parameters reach $L/2$.
    \item The lists containing the intra-cell and inter-cell hopping parameters are then doubled so that each of the remaining hopping parameters appears twice in the lists. Due to this doubling of the lists, the number of intra-cell and inter-cell hopping parameters again becomes $L$, while the mean and standard deviation of the lists remain identical as before the doubling. The two lists are then randomly shuffled.
    \item We iterate decimation and doubling processes mentioned in points 1, 2, 3, 4.
    After $N$ iterations (decimation and doubling processes), we calculate the mean ${\overline{v_{\rm{eff}}}}$ and ${\overline{w_{\rm{eff}}}}$ of the lists containing the remaining intra-cell and inter-cell hopping parameters. Variation of $|{\overline{v_{\rm{eff}}}}|$ and $|{\overline{w_{\rm{eff}}}}|$ with the number of iterations $N$ are provided in supplemental material (SM)~\cite{sm_sdrg}.  We find that for $N \gtrsim 10$, the ratio $|{\overline{w_{\rm{eff}}}}/{\overline{v_{\rm{eff}}}}|$ approaches zero or infinity depending on the value of $\lambda$, as we discuss next.
    \end{enumerate}

    Interestingly the limit of $|{\overline{w_{\rm{eff}}}}/{\overline{v_{\rm{eff}}}}| \rightarrow 0$ corresponds to an atomic insulator with just intra-cell hoppings and $|{\overline{w_{\rm{eff}}}}/{\overline{v_{\rm{eff}}}}| \rightarrow \infty$ corresponds to the other atomic insulator limit of SSH chain where just the inter-cell hoppings are present.  When $0<\lambda<1$, we find that $|{\overline{w_{\rm{eff}}}}|>|{\overline{v_{\rm{eff}}}}|$ for large $\sigma_{w}$ and all ${\overline{w}}/{\overline{v}}$ (see Fig.~\ref{fig_sdrg}(c)). However, when $\lambda>1$, the situation is different, where large $\sigma_{w}$ leads to $|{\overline{w_{\rm{eff}}}}|<|{\overline{v_{\rm{eff}}}}|$ for all ${\overline{w}}/{\overline{v}}$ (see Fig.~\ref{fig_sdrg}(d)). To analyze this contrasting behavior of $|{\overline{w_{\rm{eff}}}}/{\overline{v_{\rm{eff}}}}|$ for \text{$\lambda<1$} and \text{$\lambda>1$} in large $\sigma_{w}$ limit, we explore polarization of the strongly disordered chain, as discussed below.

\begin{figure}
    \centering
\includegraphics[width=1.0\linewidth]{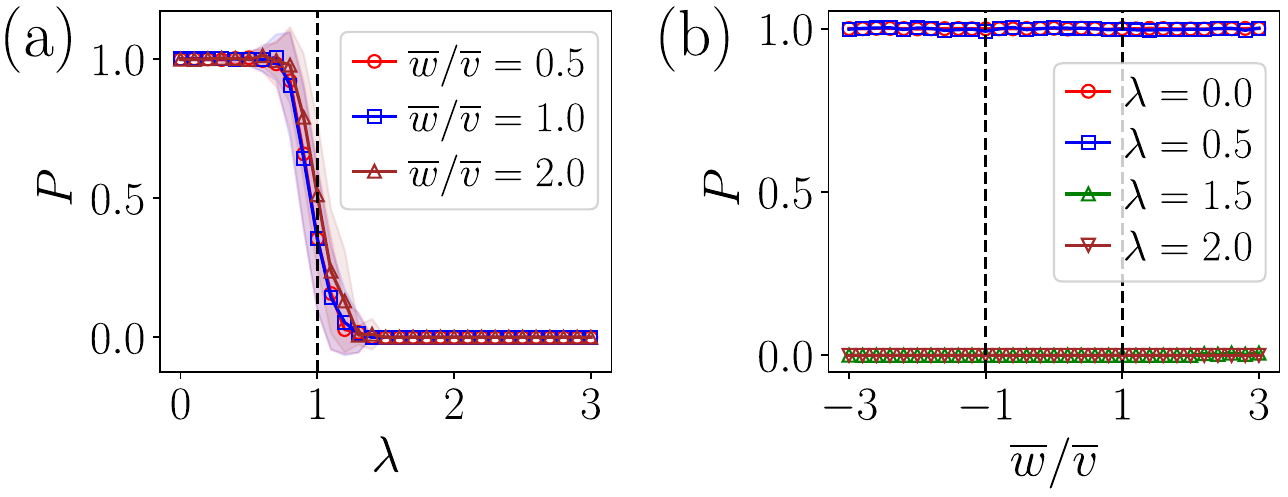} 
\caption{{\textbf{Polarization for strongly disordered chain:}} (a) 
Polarization $P$ as a function of $\lambda$ for various ${\overline{w}}/{\overline{v}}$, (b) $P$ as a function of ${\overline{w}}/{\overline{v}}$ for various $\lambda$. In all plots, $\sigma_{w}=6.0$, $L=200$ and number of realizations considered is $200$. Transition from $P=1$ to $P=0$ occurs at $\lambda= 1$ for all ${\overline{w}}/{\overline{v}}$.}
\label{fig_pol}
\end{figure}

{\it Polarization of strongly disordered chain:} We numerically compute polarization ($P$)~\cite{resta_1998} averaged over various realizations of strongly disordered chain with large $\sigma_{w}$ (see SM~\cite{sm_sdrg}). Here, we find that when $\lambda<1$, $P \sim 1$, thus confirming a topological phase, while the chain is trivial and $P \sim 0$ when $\lambda>1$ (see Fig.~\ref{fig_pol}(a)). Furthermore, $P$ does not change with ${\overline{w}}/{\overline{v}}$ for both $\lambda<1$ and $\lambda>1$, as can be seen from Fig.~\ref{fig_pol}(b). This establishes that the quantity $|{\overline{w_{\rm{eff}}}}/{\overline{v_{\rm{eff}}}}|$ calculated using SDRG methods is consistent with the numerical study of $P$, where $|{\overline{w_{\rm{eff}}}}|>|{\overline{v_{\rm{eff}}}}|$ corresponds to topological phase ($P \sim 1$), while $|{\overline{w_{\rm{eff}}}}|<|{\overline{v_{\rm{eff}}}}|$ implies a trivial phase ($P \sim 0$). Therefore, the transition from topological phase to trivial phase at $\lambda = 1$ is captured by $P$, regardless of the value of ${\overline{w}}/{\overline{v}}$. We next inquire into energy-gap and zero modes in these phases.

\begin{figure}[hbt!]
    \centering
\includegraphics[width=1.0\linewidth]{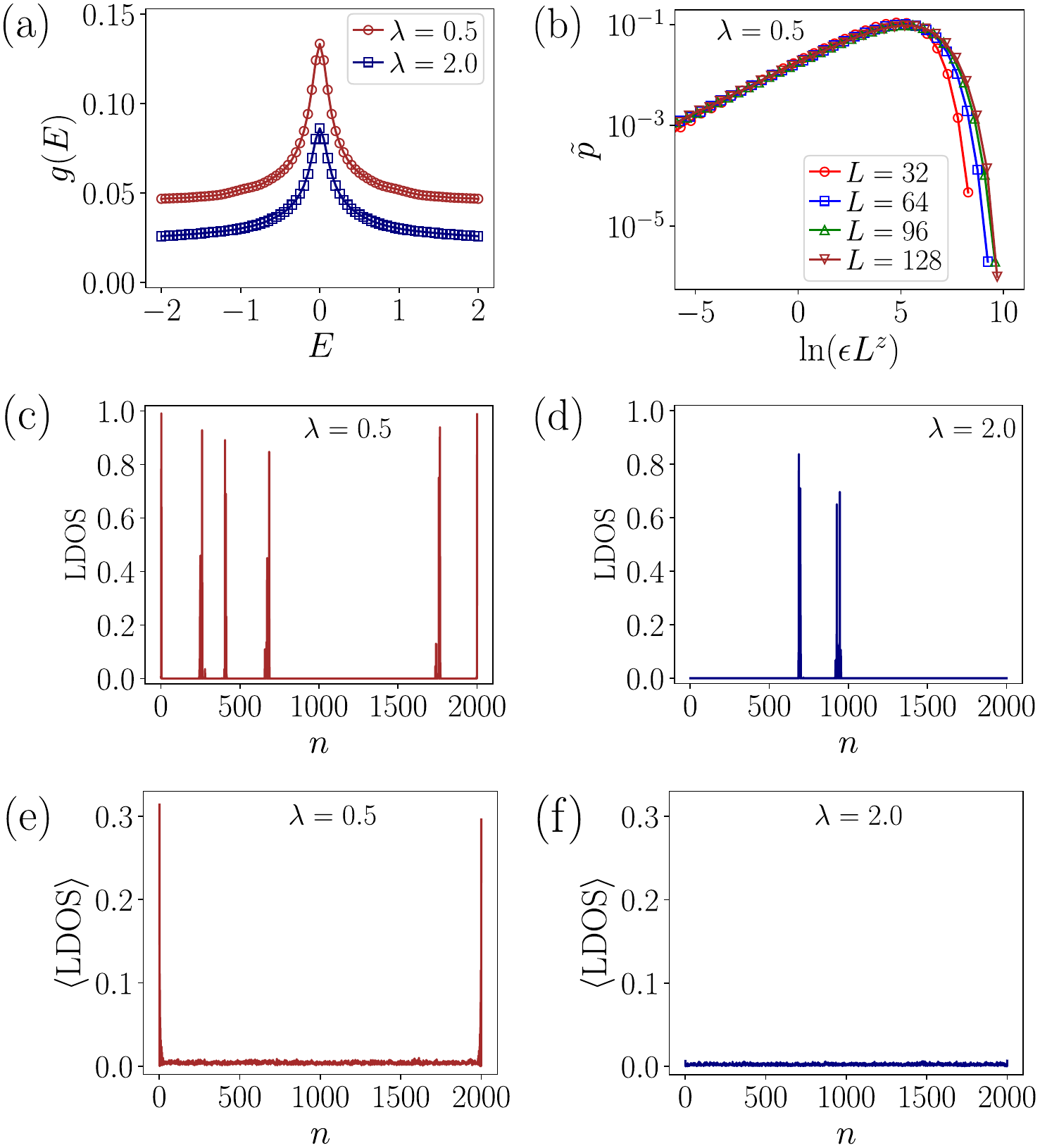}
\caption{{\textbf{Energy gap and edge modes:}} (a) Density of states $g(E)$ as a function of energy $E$ for system size $L=1000$ where averaging is performed over $1000$ realizations of disordered configurations. Finite $g(E)$ at $E \to 0$ suggests gapless phase for $\lambda<1$ and $\lambda>1$. (b) Probability $\tilde{p}$ of having bulk energy gap $\epsilon$ in periodic boundary condition as a function of $\ln(\epsilon L^{z})$ for $\lambda=0.5$. We have considered $10^{6}$ realizations and collapse of $\tilde{p}$ for various $L$ with $z=2.1$ confirms Griffiths scaling of $\epsilon$ (see Eq.~\eqref{eq_griffiths}). (c,d) Local density of states (LDOS) of zero modes (with single-particle energy eigenvalues $|E_{j}|<10^{-3}$) as a function of site index $n$ in a typical configuration with $L=1000$ in open boundary condition (where number of sites is $2000$) for (c) $\lambda=0.5$, (d) $\lambda=2.0$, where we observe localized zero modes even within bulk. (e,f) LDOS averaged over $1000$ realizations ($\langle {\rm{LDOS}} \rangle$) with $L=1000$ in open boundary condition as a function of $n$ for (e) $\lambda=0.5$, (f) $\lambda=2.0$. While $\langle {\rm{LDOS}} \rangle$ reveals zero modes localized at two edges for $\lambda<1$, there is no such edge-localized zero modes for $\lambda>1$. In all plots, ${\overline{w}}/{\overline{v}}=0.5$, $\sigma_{w}=6.0$.}
\label{fig_gap_ldos}
\end{figure}

{\it Energy-gap and zero modes:} We analyze the single-particle spectrum of disordered chain and find that density of states $g(E)$ is non-zero at $E\to 0$ for both $\lambda<1$ and $\lambda>1$ (see Fig.~\ref{fig_gap_ldos}(a)), indicating that topological ($\lambda<1$) and trivial ($\lambda>1$) phases are gapless. Furthermore, bulk energy-gap $\epsilon$ for a finite system size $L$ in periodic boundary condition shows Griffiths scaling:
\begin{equation}\label{eq_griffiths}
    \epsilon \sim L^{-z},
\end{equation}
with dynamical exponent $z>0$~\cite{young_prb_1996}, which we confirm from the collapse of the probability $\tilde{p}$ of having bulk energy-gap $\epsilon$ for various $L$ while plotted against $\ln(\epsilon L^{z})$ (see Fig.~\ref{fig_gap_ldos}(b)). In fact, $z$ is found to decrease as a power-law with $|\lambda-1|$ when $\lambda<1$ and $\lambda>1$, indicating the divergence of $z$ at $\lambda \to 1$ (see SM~\cite{sm_sdrg}). 

We then compute the local density of states ${\rm{LDOS}}(n)=\sum_{j} |\langle n | \psi_{j} \rangle |^{2}$ (where $j$ runs over single-particle energy eigenvalues $|E_{j}|<10^{-3}$) as a function of site index $n$ (where odd $n$ corresponds to $A$ sites and even $n$ corresponds to $B$ sites) in a typical configuration with open boundary conditions and find that some zero modes are localized even within the bulk of the chain when $\lambda<1$ and $\lambda>1$ (see Fig.~\ref{fig_gap_ldos}(c,d)). However, the site indices where these zero modes are localized within the bulk vary for different configurations, and thus the configuration-averaged $\langle {\rm{LDOS}} \rangle$ approaches zero for these bulk sites. Interestingly, $\langle {\rm{LDOS}} \rangle$ captures two edge-localized zero modes in topological phase ($\lambda<1$), while there is no such edge-localized zero mode in trivial phase $\lambda>1$ (see Fig.~\ref{fig_gap_ldos}(e,f)). We now proceed to analyze the properties of the QCP at $\lambda=1$, as discussed next.

\begin{figure}
    \centering
\includegraphics[width=1.0\linewidth]{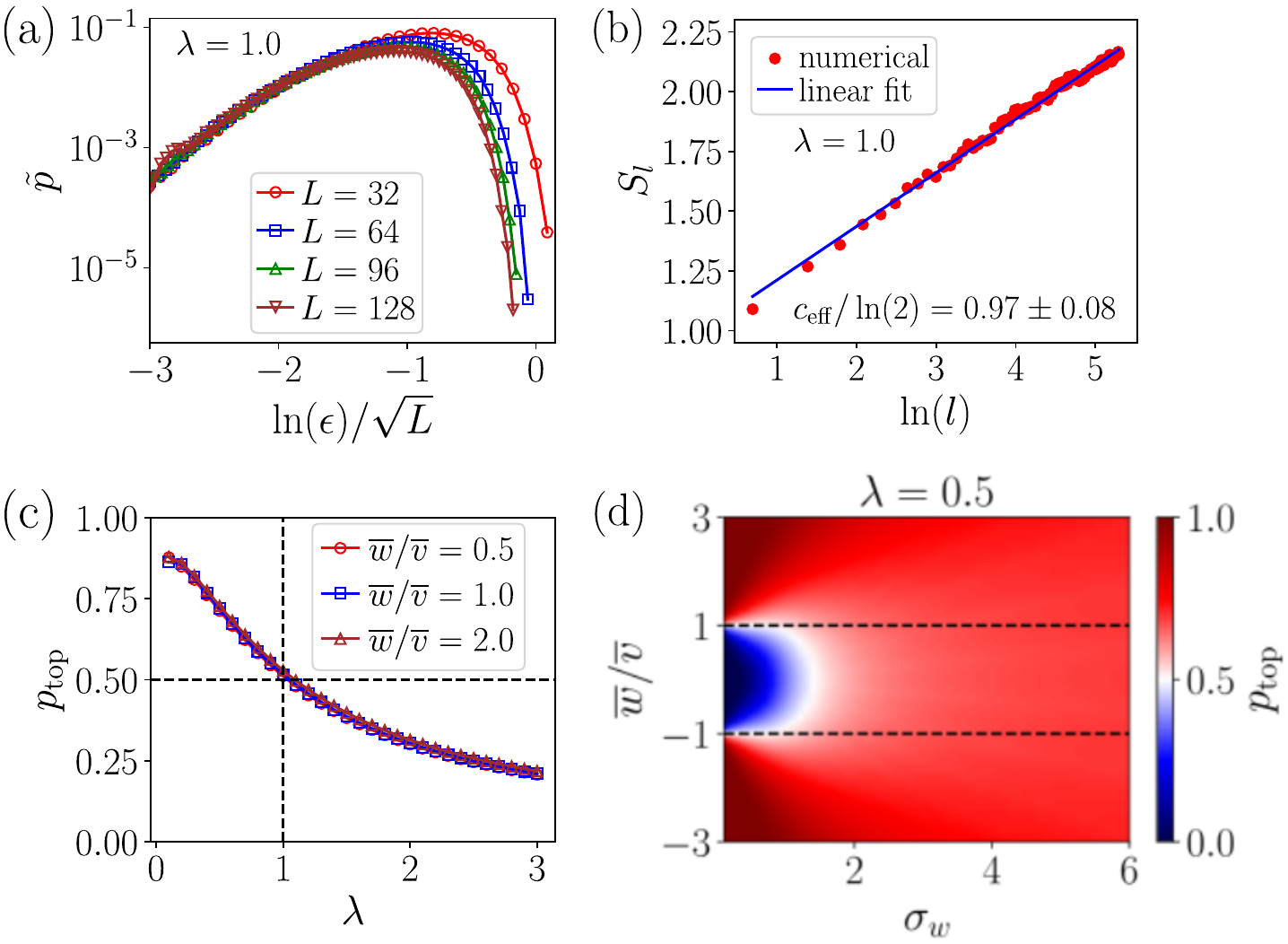}
\caption{{\textbf{Infinite randomness fixed point:}}
(a) Probability $\tilde{p}$ of having bulk energy gap $\epsilon$ as a function of $\ln(\epsilon)/{\sqrt{L}}$ when $\lambda=1.0$, ${\overline{w}}/{\overline{v}}=0.5$, $\sigma_{w}=6.0$. We have considered $10^{6}$ realizations of disordered configurations. Collapse of $\tilde{p}$ for various $L$ confirms activated scaling of $\epsilon$ (see Eq.~\eqref{eq_activated_scaling}) at $\lambda=1.0$. (b) Entanglement entropy $S_{l}$ between subsystems of size $l$ and $(L-l)$ as a function of $\ln(l)$ in periodic boundary conditions, where $L=1000$, ${\overline{w}}/{\overline{v}}=0.5$, $\sigma_{w}=6.0$, $\lambda=1.0$. $S_{l}$ is averaged over $1000$ realizations. Effective central charge $c_{\rm{eff}}$ is obtained from slope of linear fit (see Eq.~\eqref{eq_ceff}) where $c_{\rm{eff}} \approx \ln(2)$, indicating that $\lambda=1.0$ is an infinite randomness fixed point. (c) Probability $p_{\rm{top}}$ of having 
a topological unit cell (see Eq.~\eqref{eq_ptop}) as a function of $\lambda$ when $\sigma_{w}=6.0$. At $\lambda=1.0$, $p_{\rm{top}}=0.5$ for any ${\overline{w}}/{\overline{v}}$. (d) $p_{\rm{top}}$ as a function of $\sigma_{w}$ and ${\overline{w}}/{\overline{v}}$ when $\lambda=0.5$, where $p_{\rm{top}}>1/2$ in topological phase and $p_{\rm{top}}<1/2$ in trivial phase. 
}
\label{fig_eec}
\end{figure}

{\it Activated scaling of bulk energy-gap at QCP:} Bulk energy-gap $\epsilon$ at QCP $\lambda=1$ exhibits stretched exponential fall of $\epsilon$ with $L$, namely activated dynamical scaling:
\begin{equation}\label{eq_activated_scaling}
    \epsilon \sim \exp(-a \sqrt{L}),
\end{equation}
where $a>0$ is a constant. To verify activated scaling at $\lambda=1$, we numerically compute the probability ${\tilde{p}}$ of having energy gap $\epsilon$, which is then plotted against $\ln(\epsilon)/{\sqrt{L}}$~\cite{young_prb_1996}. Collapse of these plots at smaller $\epsilon$ for various $L$ confirms activated dynamical scaling of $\epsilon$ at QCP $\lambda=1$ (see Fig.~\ref{fig_eec}(a)). We next move on to discuss central charge associated with the QCP.

{\it Irrational central charge:}
In order to calculate central charge at QCP $\lambda=1$, we investigate the scaling of entanglement entropy $S_{l}$ between two subsystems with $l$ and $(L-l)$ unit cells in a strongly disordered chain~\cite{vidal_prl_2003,calabrese_jsmte_2004,latorre_jpamt_2009,calabrese_jpamt_2009}. When $l \ll L$, we find characteristic logarithmic dependence of $S_{l}$ (see Fig.~\ref{fig_eec}(b)), leading to scaling relation:
\begin{equation}\label{eq_ceff}
S_{l}= \frac{c_{\rm{eff}}}{3} \ln(l) + c_{0},
\end{equation}
where $c_{\rm{eff}}$ is effective central charge and $c_{0}$ is a non-universal constant. Using linear fit of $S_{l}$ with $\ln(l)$ as given in Eq.~\eqref{eq_ceff}, we surprisingly find that $c_{\rm{eff}} \approx \ln(2)$ at $\lambda = 1$, which is $\ln(2)$ times the central charge $c=1$ associated with the QCPs $w=\pm v$ for a clean SSH chain. Irrational central charge along with activated dynamical scaling establishes that the QCP at $\lambda=1$ is an IRFP. Interestingly, the IRFP at $\lambda=1$ occurs for any ${\overline{w}}/{\overline{v}}$ (see SM~\cite{sm_sdrg}). We next explore the determination of QCP from probability distributions of hopping parameters.

{\it Determination of QCP from probability distributions:}
We calculate the probability $p_{\rm{top}}$ of having a topological unit cell using normal probability distributions $p_{1}(v)$ and $p_{2}(w)$ of intra-cell hopping $v$ and inter-cell hopping $w$ (see SM~\cite{sm_sdrg}). Here, a unit cell in the disordered chain can be topological if the associated $w$ and $v$ satisfy $|w|>|v|$ and thus,
\begin{equation}\label{eq_ptop}
    p_{\rm{top}}= \int_{-\infty}^{\infty} dw p_{2}(w) \int_{-|w|}^{|w|} dv p_{1}(v).
\end{equation}
Performing numerical integration, we find that for large $\sigma_{w}$, \text{$p_{\rm{top}}>1/2$} in topological phase (\text{$\lambda<1$}) and \text{$p_{\rm{top}}<1/2$} in trivial phase (\text{$\lambda>1$}). Interestingly, \text{$p_{\rm{top}} \sim 1/2$} at \text{$\lambda = 1$} for all \text{${\overline{w}}/{\overline{v}}$}, thus confirming topological phase transition at \text{$\lambda = 1$} under strong disorder (see Fig.~\ref{fig_eec}(c)). Furthermore, when \text{$\lambda<1$} (\text{$\lambda>1$}) and \text{${|\overline{w}}|<{|\overline{v}|}$} (\text{${|\overline{w}}|>{|\overline{v}|}$}), tuning $\sigma_{w}$ leads to a transition from trivial (topological) to topological (trivial) phase (see SM~\cite{sm_sdrg}). This transition is also captured by $p_{\rm{top}}$, as shown in Fig.~\ref{fig_eec}(d) which is consistent with the SDRG results as shown in Fig.~\ref{fig_sdrg}(c). 

While we have analyzed strong disorder-driven topological phase transitions for normal probability distribution of hopping parameters, the results remain qualitatively the same when hopping parameters are chosen from other probability distributions, such as uniform probability distribution (see SM~\cite{sm_sdrg}).

{\it Outlook:} In this work, we explore topological phase transitions induced by strong randomness. Considering strongly disordered SSH chain, we here show that transition from topological to trivial phase is possible by tuning the ratio of disorder energy scales of the intra-cell and inter-cell hopping parameters. Interestingly, the QCP shows activated dynamical scaling and irrational central charge, consistent with the IRFP. Our work opens up future questions regarding the exploration of the nature of topological phases and transitions in strongly disordered systems with long-range couplings and higher-dimensional disordered systems (e.g. disordered Chern insulator) using SDRG methods. In topological insulators and superconductors, where multiple parameters are random and chosen from various probability distributions, interplay of infinite randomness and multicriticality can also be studied. In addition, investigating topological and entanglement properties of strongly disordered systems in non-equilibrium situations such as sudden quench, periodic driving is another interesting future question~\cite{vasseur_prb_2018}. 

{\it Acknowledgements:}
We acknowledge fruitful discussions with Diptiman Sen. S.M. acknowledges support from PMRF Fellowship, India. AA acknowledges the kind hospitality of ICTS via the ICTS Associates Program. Numerical calculations were performed on the workstation {\it Wigner} at IIT Kanpur.

\bibliography{reference.bib}

%\documentclass[aps,prx,floats,epsfig,twocolumn]{revtex4-2}

% \usepackage{graphicx}
% \usepackage{amsmath}
% \usepackage{amssymb}
% \usepackage{braket}
% \usepackage{bbold}
% \usepackage{hyperref}
% \usepackage{subfigure}
% \usepackage[usenames,dvipsnames]{color}
% \usepackage{float}
% \usepackage{graphicx}
% \usepackage{subfigure}
% \usepackage[mathscr]{euscript}
% \hypersetup{colorlinks,allcolors=black}

%  \newcommand{\Tr}{\rm{Tr}}
%  \newcommand{\im}{\rm{Im}}
%  \newcommand{\real}{\rm{Re}}
%  \newcommand{\Li}{\rm{Li}}
%  \newcommand{\sign}{\operatorname{\rm{sign}}}
%  \newcommand{\tb}{\textcolor{black}}
%   \newcommand{\tbb}{\textcolor{blue}}
%  \newcommand{\trr}{\textcolor{black}}
%  \newcommand{\trrr}{\textcolor{black}}

%  \begin{document}
%  \title{Supplemental Material to ``Topological Phase Transition under Infinite Randomness"}

 % \author{Saikat Mondal}
 % \email{msaikat@iitk.ac.in}
 % \affiliation{Department of Physics, Indian Institute of Technology Kanpur, Kalyanpur, Uttar Pradesh 208016, India}

 % \author{Adhip Agarwala}
 % \email{adhip@iitk.ac.in}
 % \affiliation{Department of Physics, Indian Institute of Technology Kanpur, Kalyanpur, Uttar Pradesh 208016, India}

 %\maketitle

%\newpage
\setcounter{equation}{0}
\setcounter{figure}{0}
\makeatletter
\renewcommand{\theequation}{S\arabic{equation}}
\renewcommand{\thefigure}{S\arabic{figure}}

 \onecolumngrid

\begin{center}
	\textbf{\large Supplemental Material to ``Topological Phase Transition under Infinite Randomness"}
\end{center}

 \vspace{\columnsep}
 %\vspace{\columnsep}

 \twocolumngrid

\section{SDRG results at various iterations}
\begin{figure}[hbt!]
    \centering
\includegraphics[width=1.0\linewidth]{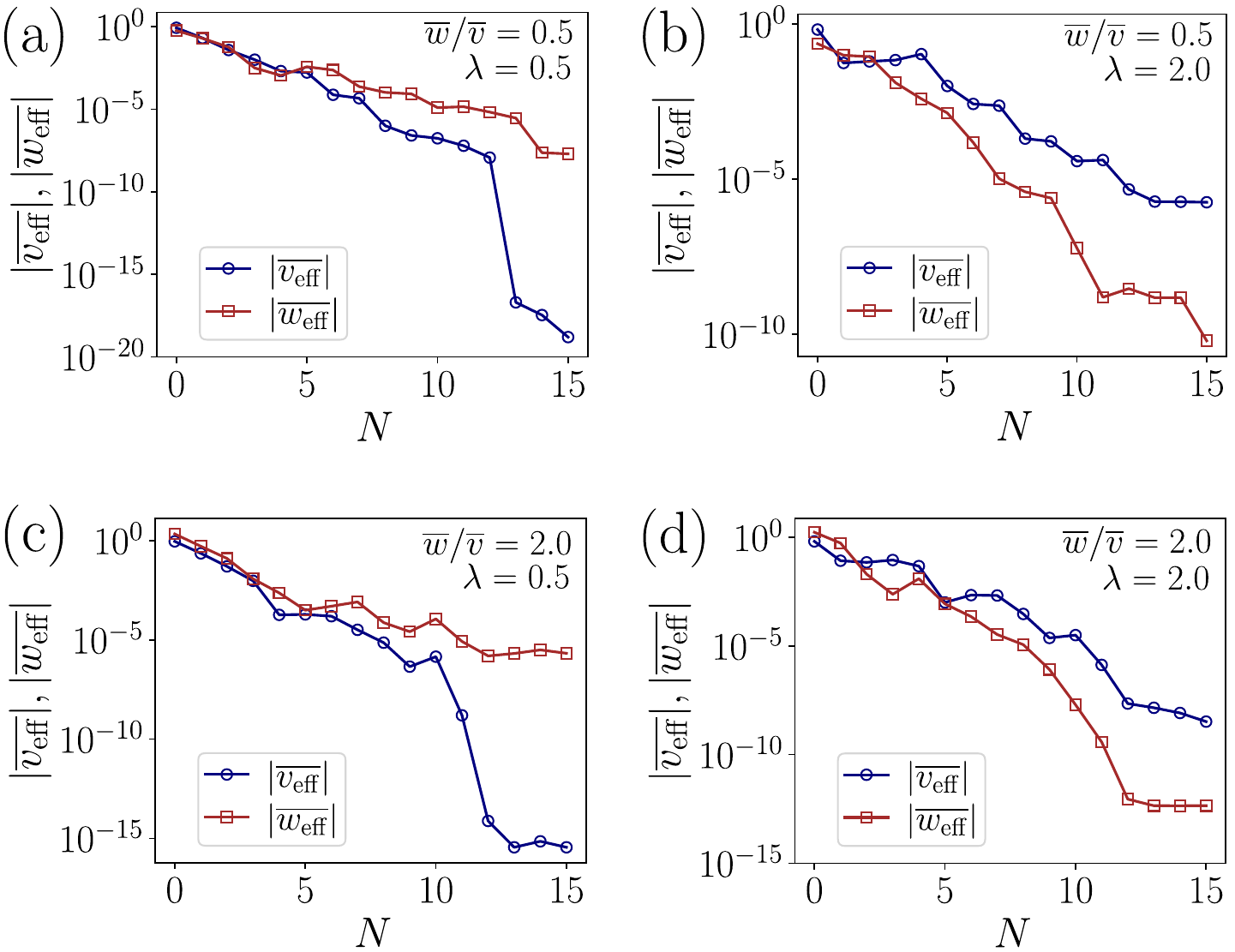}
\caption{{\textbf{SDRG results at various iterations:}}  $|{\overline{v_{\rm{eff}}}}|$ and $|{\overline{w_{\rm{eff}}}}|$ as a function of the number of iterations $N$ in SDRG methods when (a) ${\overline{w}}/{\overline{v}}=0.5$, $\lambda=0.5$, (b) ${\overline{w}}/{\overline{v}}=0.5$, $\lambda=2.0$, (c) ${\overline{w}}/{\overline{v}}=2.0$, $\lambda=0.5$, (d) ${\overline{w}}/{\overline{v}}=2.0$, $\lambda=2.0$. In all plots, $L=1000$, $\sigma_{w}=6.0$. For $N \gtrsim 10$, we find $|{\overline{w_{\rm{eff}}}}|>|{\overline{v_{\rm{eff}}}}|$ when $\lambda<1$ and $|{\overline{w_{\rm{eff}}}}|<|{\overline{v_{\rm{eff}}}}|$ when $\lambda>1$, regardless of the value of ${\overline{w}}/{\overline{v}}$.
}
\label{fig_sdrg_steps}
\end{figure}

In this section, we analyze the SDRG results of strongly disordered SSH chain for different numbers of iterations. After each iteration, we calculate the means ${\overline{v_{\rm{eff}}}}$ and ${\overline{w_{\rm{eff}}}}$ of the two lists containing the effective intra-cell and inter-cell hopping parameters. We then plot $|{\overline{v_{\rm{eff}}}}|$ and $|{\overline{w_{\rm{eff}}}}|$ as a function of the number of iterations $N$ for various initial values of ${\overline{w}}/{\overline{v}}$ and $\lambda=\sigma_{v}/\sigma_{w}$. Here, we find that for $N \gtrsim$ 10, $|{\overline{w_{\rm{eff}}}}|>|{\overline{v_{\rm{eff}}}}|$ when $\lambda<1$ and $|{\overline{w_{\rm{eff}}}}|<|{\overline{v_{\rm{eff}}}}|$ when $\lambda>1$, regardless of the value of ${\overline{w}}/{\overline{v}}$ (see Fig.~\ref{fig_sdrg_steps}(a-d)).

\section{Polarization}
We here discuss the method used to compute polarization of a strongly disordered SSH chain. First, we define the position operator in a SSH chain
\begin{equation}
    \hat{X}=\sum_{n=1}^{L} n (c_{n,A}^{\dagger} c_{n,A} + c_{n,B}^{\dagger} c_{n,B}).
\end{equation}
Single-particle energy eigenvalues and the corresponding eigenvectors of Hamiltonian in Eq.~(1) of main text are then obtained. For a SSH chain with $L$ unit cells, there are $2L$ single-particle energy levels which are arranged in ascending order of energies and indexed as $j=1,2,...,2L$. If $| \psi_{j} \rangle $ is the eigenvector corresponding to $j$-th energy eigenvalue, then the projection operator at half-filling is given by
\begin{equation}
    \hat{{\mathcal{P}}}= \sum_{j=1}^{L} |\psi_{j} \rangle \langle \psi_{j}|.
\end{equation}
The polarization $P$ is then obtained using
\begin{equation}\label{eq_pol}
P = {\rm{Im}} \left[ \frac{1}{\pi} {\rm{Tr}} \left[ \ln \left(\hat{{\mathcal{P}}} \exp \left(i \frac{2\pi \hat{X}}{L} \right) \hat{{\mathcal{P}}} \right) \right] \right] \text{~~modulo $(2)$}.
\end{equation}
In a disordered SSH chain, polarization for various realizations of disordered configurations are computed and averaged.

\section{Bulk energy gap and entanglement entropy}
\begin{figure}
    \centering
\includegraphics[width=1.0\linewidth]{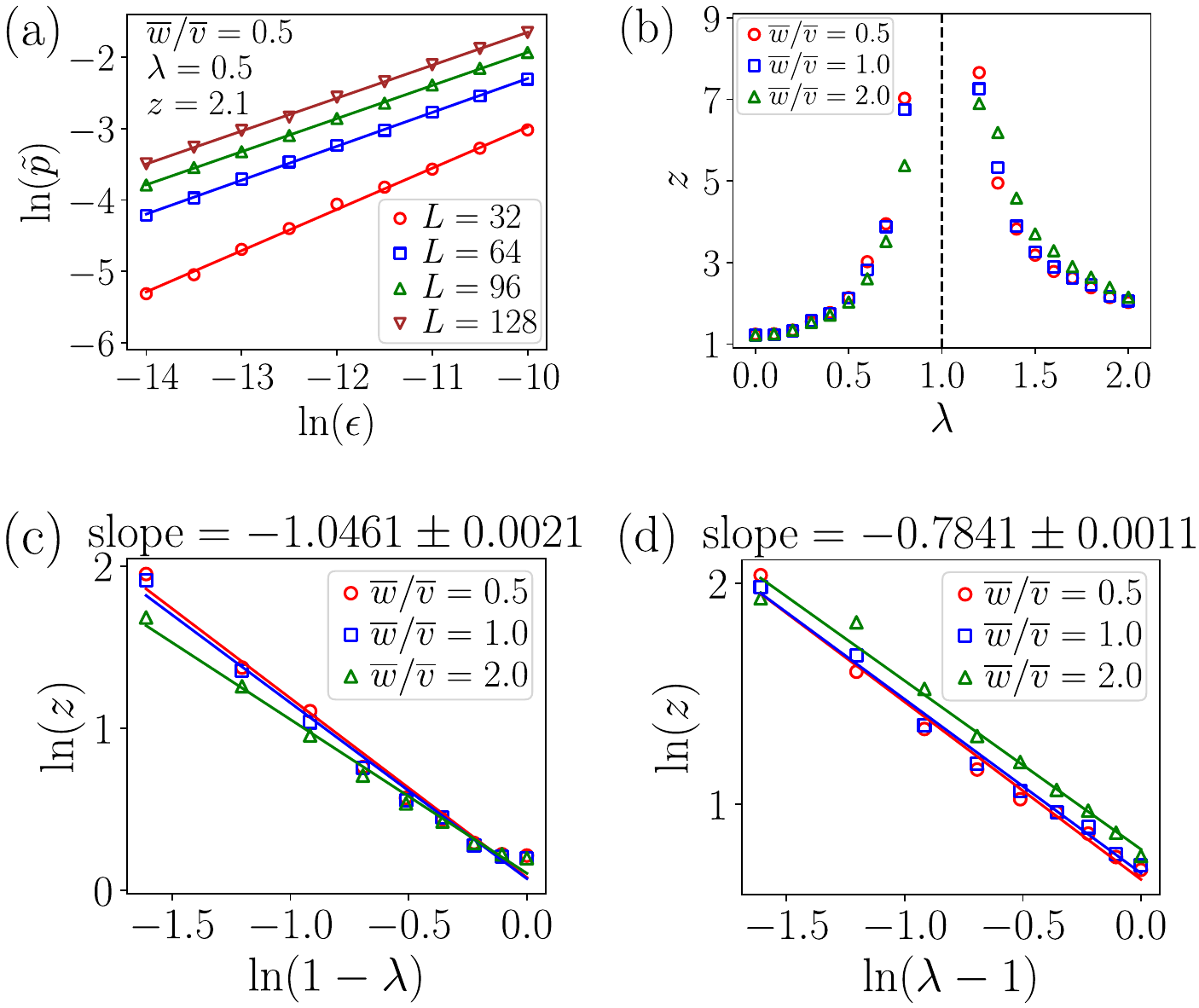}
\caption{{\textbf{Griffiths scaling and $z$ exponent:}} (a) $\ln({\tilde{p}})$ as a function of $\ln(\epsilon)$ where ${\tilde{p}}$ is the probability of having bulk energy-gap $\epsilon$ in periodic boundary conditions at $\lambda=0.5$, ${\overline{w}}/{\overline{v}}=0.5$, $\sigma_{w}=6.0$. Number of realizations considered is $10^{6}$. From the slope of linear fit using Eq.~\eqref{eq_pz}, we obtain $z=2.1$. (b) $z$ exponent as a function of $\lambda$, where diverging behavior of $z$ at $\lambda \to 1$ is indicated. (c) $\ln(z)$ as a function of $\ln(1-\lambda)$ for $\lambda<1$ where slope of linear fit is $-1.0461 \pm 0.0021$. (d) $\ln(z)$ as a function of $\ln(\lambda-1)$ for $\lambda>1$ where slope of linear fit is $-0.7841 \pm 0.0011$. These observations confirm that $z$ falls as a power-law with $|\lambda-1|$ when $\lambda<1$ and $\lambda>1$. 
}
\label{fig_z}
\end{figure}

\begin{figure}
    \centering
\includegraphics[width=1.0\linewidth]{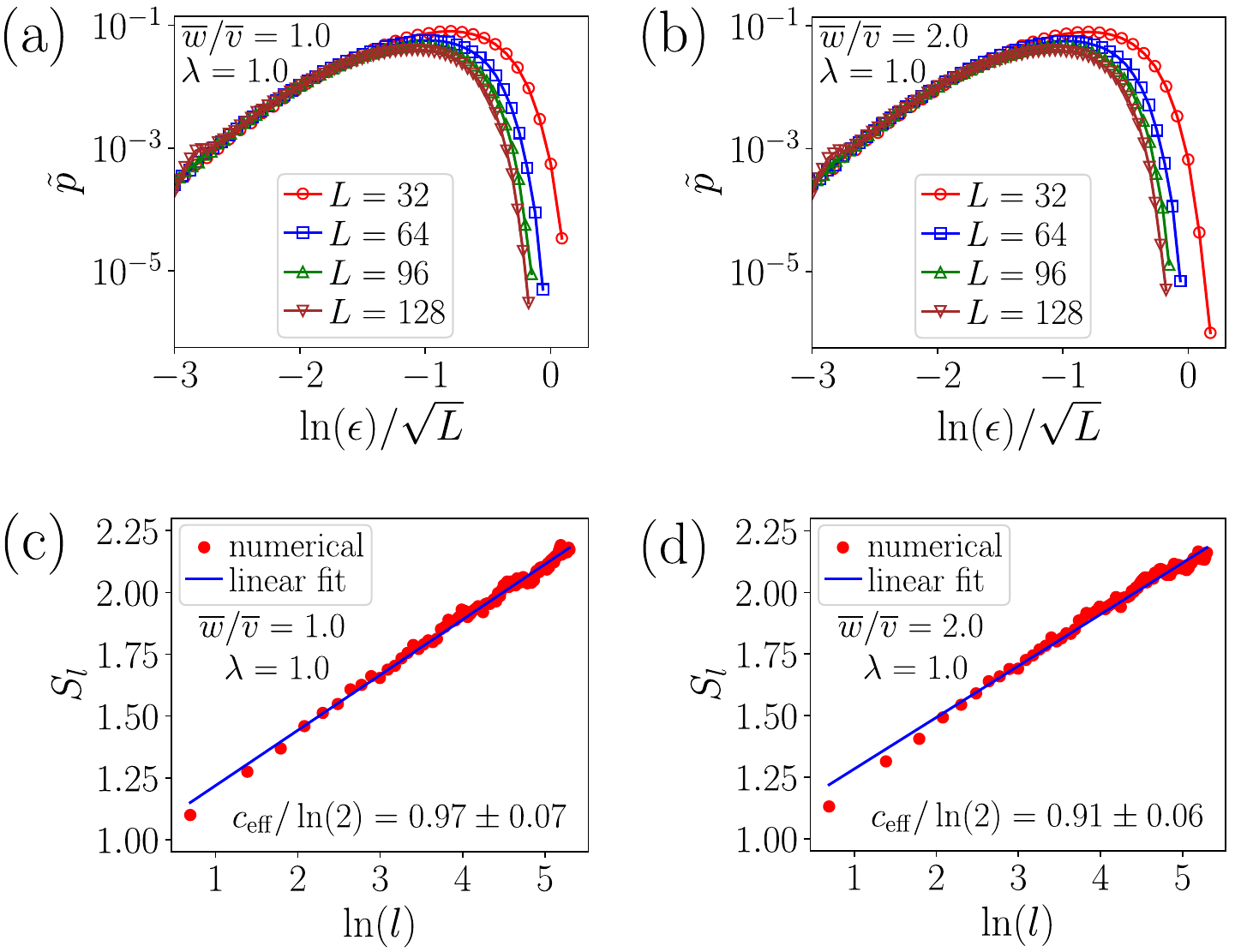}
\caption{{\textbf{Activated scaling of energy-gap and irrational central charge:}} (a,b) Probability ${\tilde{p}}$ of having bulk energy-gap $\epsilon$ in periodic boundary conditions at $\lambda=1.0$ as a function of $\ln(\epsilon)/{\sqrt{L}}$ when (a) ${\overline{w}}/{\overline{v}}=1.0$, (b) ${\overline{w}}/{\overline{v}}=2.0$. In both the plots, $\sigma_{w}=6.0$ and $10^{6}$ realizations of disordered configurations have been considered. Collapse for various $L$ confirms activated scaling of energy-gap at $\lambda=1.0$, irrespective of the value of ${\overline{w}}/{\overline{v}}$. (c,d) Entanglement entropy $S_{l}$ between subsystems of size $l$ and $(L-l)$ as a function of $\ln(l)$ in periodic boundary conditions at $\lambda=1.0$ when (c) ${\overline{w}}/{\overline{v}}=1.0$, (d) ${\overline{w}}/{\overline{v}}=2.0$. In both the plots, $\sigma_{w}=6.0$, $L=1000$ and $S_{l}$ is averaged over $1000$ realizations. We find $c_{\rm{eff}} \approx \ln(2)$ for all ${\overline{w}}/{\overline{v}}$.
}
\label{fig_c}
\end{figure}

\begin{figure*}
    \centering
\includegraphics[width=0.9\linewidth]{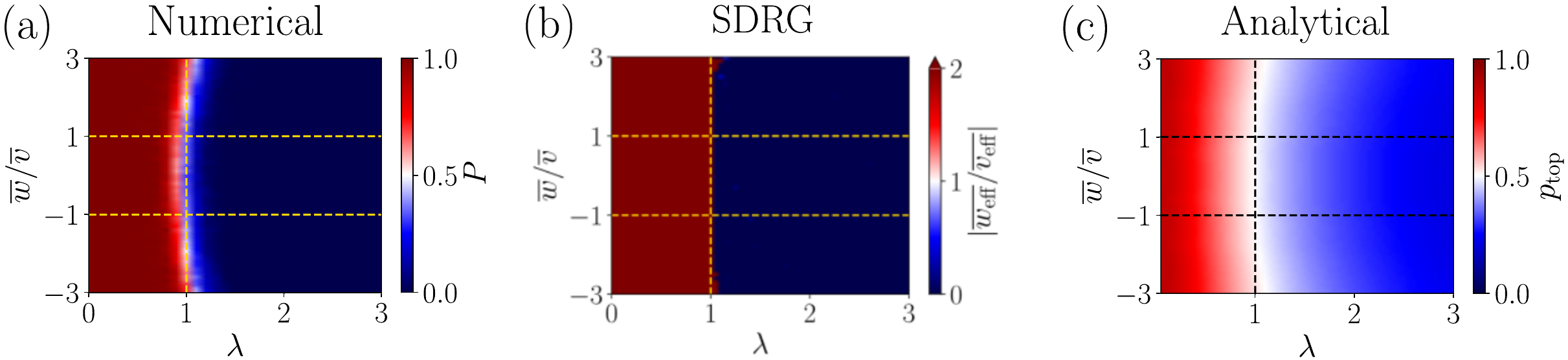}
\caption{{\textbf{Comparison of numerical, SDRG and analytical methods:}} (a) Polarization $P$ as a function of ${\overline{w}}/{\overline{v}}$ and $\lambda$, (b) $|\overline{w_{\rm{eff}}}/\overline{v_{\rm{eff}}}|$ calculated using SDRG methods for the number of iterations $N=15$ as a function of ${\overline{w}}/{\overline{v}}$ and $\lambda$, (c) Probability $p_{\rm{top}}$ of having 
a topological unit cell (see Eq.~\eqref{eq_ptop_detail}) as a function of ${\overline{w}}/{\overline{v}}$ and $\lambda$. In all plots, $\sigma_{w}=6.0$. $P$ is averaged over $200$ realizations in (a) and $100$ realizations are considered in (b). We find that $p_{\rm{top}} \sim 1/2$ at the quantum critical point $\lambda=1$ for any value of ${\overline{w}}/{\overline{v}}$.
}
\label{fig_comparison}
\end{figure*}

In this section, we discuss the methods used to compute bulk energy gap and entanglement entropy along with additional numerical results.

For the computation of bulk energy-gap, we resort to periodic boundary conditions and find the single-particle energy-eigenvalues $E_{j}$ (where $j=1,2,...,2L$ and these are arranged in ascending order of energy) for each realization of disordered configurations. Then, the bulk energy gap for each of such realizations is calculated as $\epsilon=E_{L+1}-E_{L}$. Considering $10^{6}$ realizations, we then obtain probability distribution $\tilde{p}$ of bulk energy-gap $\epsilon$. For Giffiths scaling of $\epsilon$, $\ln(\tilde{p})$ is known to show the following variation with $\ln(\epsilon)$:
\begin{equation}\label{eq_pz}
    \ln(\tilde{p})=\frac{1}{z} \ln(\epsilon) + {\rm{constant}},
\end{equation}
where dynamical exponent $z$ is obtained from the slope of linear fit of $\ln(\tilde{p})$ with $\ln(\epsilon)$ (see Ref.~[27]). To verify Griffiths scaling in topological (\text{$\lambda<1$}) and trivial (\text{$\lambda>1$}) phases, $\ln(\tilde{p})$ is plotted against $\ln(\epsilon)$. We find that $\ln(\tilde{p})$ varies linearly with $\ln(\epsilon)$ and thus $z=1/{\rm{slope}}$ (see Fig.~\ref{fig_z}(a)). The variation of $z$ exponent with $\lambda$ is shown in Fig.~\ref{fig_z}(b), where $z$ increases as $\lambda=1$ is approached from above and below. This indicates divergence of the effective $z$ at $\lambda \to 1$. Interestingly, $z$ falls as a power law with $|\lambda-1|$ when $\lambda<1$ and $\lambda>1$, as evident from linear dependence of $\ln(z)$ on $\ln(|\lambda-1|)$ (see Fig.~\ref{fig_z}(c,d)). Now, we move on to discuss activated scaling of $\epsilon$ at $\lambda=1$, as discussed below.

To confirm that activated scaling at $\lambda=1$ occurs regardless of the value of ${\overline{w}}/{\overline{v}}$, the probability distribution $\tilde{p}$ for $\lambda=1$ is plotted against $\ln(\epsilon)/\sqrt{L}$ when ${\overline{w}}/{\overline{v}}=1.0$ and ${\overline{w}}/{\overline{v}}=2.0$ (see Fig.~\ref{fig_c}(a,b)). For both the values of ${\overline{w}}/{\overline{v}}$, collapse of $\tilde{p}$ for various $L$ confirms activated scaling $\epsilon \sim \exp(-a \sqrt{L})$ (where $a={\rm{constant}}$)  at QCP $\lambda=1.0$, irrespective of the value of ${\overline{w}}/{\overline{v}}$.

To calculate the entanglement entropy, we first evaluate the $2L \times 2L$ correlation matrix $C$ whose elements are $C_{mn}=\langle c_{m}^{\dagger} c_{n} \rangle$ (where $m,n=1,...,2L$ are site indices). At half-filling, we thus calculate
\begin{equation}
    C_{mn}=\langle c_{m}^{\dagger} c_{n} \rangle= \sum_{j=1}^{L} \langle \psi_{j}| c_{m}^{\dagger} c_{n} | \psi_{j} \rangle.
\end{equation}
Now, entanglement entropy $S_{l}$ between two subsystems having $l$ and $(L-l)$ unit cells is obtained from eigenvalues ($\lambda_{j}$) of the $2 l \times 2l$ reduced correlation matrix $C^{\prime}$ having the elements $C^{\prime}_{mn}=C_{mn}$ (where $m,n=1,...,2l$) as:
\begin{equation}
    S_{l} =\sum_{j=1}^{2l} \left( -\lambda_{j} \ln(\lambda_{j})-(1-\lambda_{j})\ln(1-\lambda_{j}) \right).
\end{equation}
We show the variation of $S_{l}$ with $\ln(l)$ for $\lambda=1.0$ when ${\overline{w}}/{\overline{v}}=1.0$ and ${\overline{w}}/{\overline{v}}=2.0$ in Fig.~\ref{fig_c}(c,d). In both the plots, we obtain a linear increase of $S_{l}$ with $\ln(l)$ and effective central charge is calculated as $c_{\rm{eff}}=3 \times {\rm{slope}}$ of the linear fit. We find that $c_{\rm{eff}} \approx \ln(2)$ at the QCP $\lambda=1.0$ for both the values of ${\overline{w}}/{\overline{v}}$.

\begin{figure*}
    \centering
\includegraphics[width=0.9\linewidth]{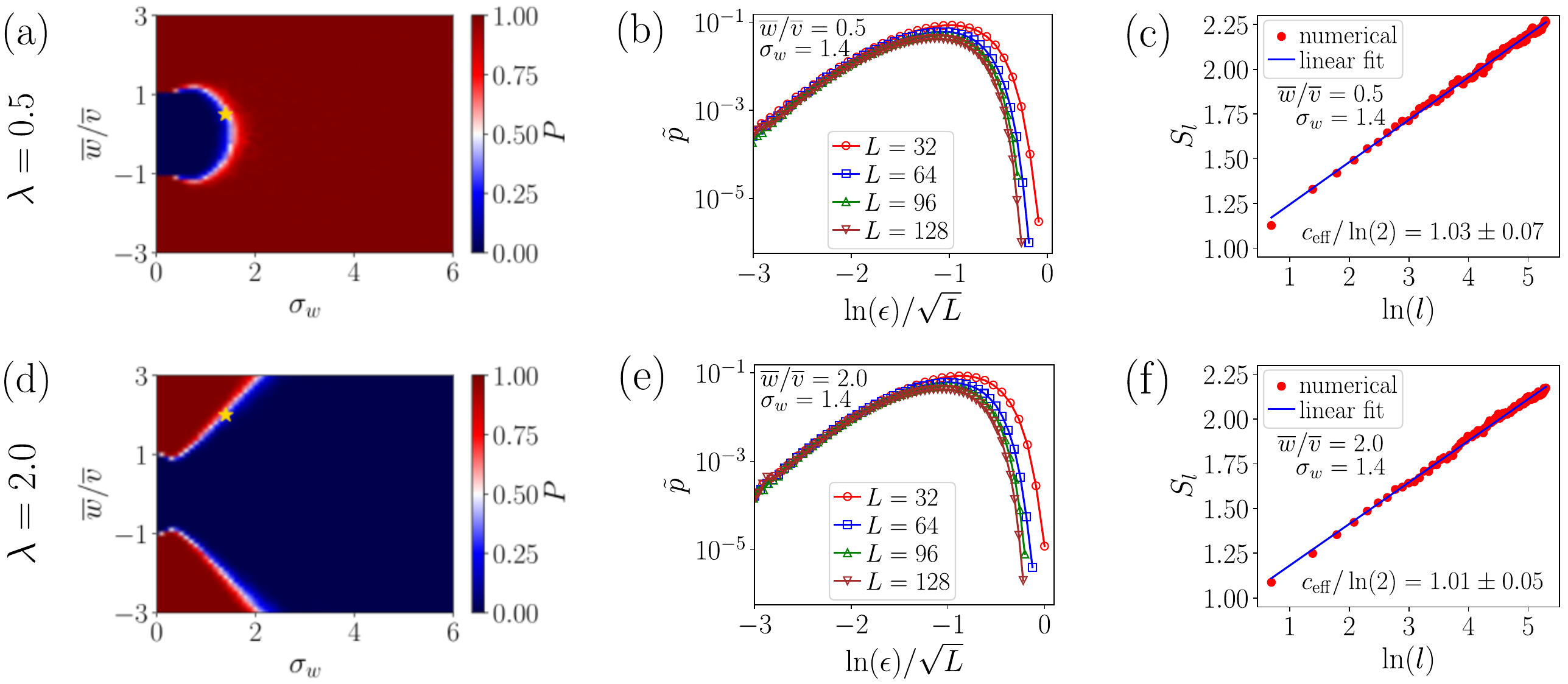}
\caption{{\textbf{$\sigma_{w}$-driven topological phase transitions:}} (a) Polarization $P$ as a function of $\sigma_{w}$ and ${\overline{w}}/{\overline{v}}$ when $\lambda=0.5$. When $|{\overline{w}}|<|{\overline{v}}|$, the transition from trivial ($P \sim 0$) to topological ($P \sim 1$) phase occurs at a finite $\sigma_{w}$ (e.g. $\sigma_{w}=1.4$ when ${\overline{w}}/{\overline{v}}=0.5$, as denoted by star). (b) Probability ${\tilde{p}}$ of having bulk energy-gap $\epsilon$ in periodic boundary conditions as a function of $\ln(\epsilon)/{\sqrt{L}}$ for the quantum critical point (QCP) at $\sigma_{w}=1.4$ when ${\overline{w}}/{\overline{v}}=0.5$, $\lambda=0.5$. (c) Entanglement entropy $S_{l}$ between subsystems of size $l$ and $(L-l)$ as a function of $\ln(l)$ in periodic boundary conditions at QCP $\sigma_{w}=1.4$ when ${\overline{w}}/{\overline{v}}=0.5$, $\lambda=0.5$. (d) Polarization $P$ as a function of $\sigma_{w}$ and ${\overline{w}}/{\overline{v}}$ when $\lambda=2.0$. When $|{\overline{w}}|>|{\overline{v}}|$, the transition from topological ($P \sim 1$) to trivial ($P \sim 0$) phase occurs at a finite $\sigma_{w}$ (e.g. $\sigma_{w}=1.4$ when ${\overline{w}}/{\overline{v}}=2.0$, as denoted by star). (e) Probability ${\tilde{p}}$ of having bulk energy-gap $\epsilon$ in periodic boundary conditions as a function of $\ln(\epsilon)/{\sqrt{L}}$ for the QCP at $\sigma_{w}=1.4$ when ${\overline{w}}/{\overline{v}}=2.0$, $\lambda=2.0$. (f) $S_{l}$ as a function of $\ln(l)$ in periodic boundary conditions at $\sigma_{w}=1.4$ when ${\overline{w}}/{\overline{v}}=2.0$, $\lambda=2.0$. In (a,d), we have taken $L=200$ and considered $200$ realizations of disordered configurations. In (b,e), $10^{6}$ realizations of disordered configurations have been considered. In (c,f), $L=1000$ and $S_{l}$ is averaged over $1000$ realizations. We observe activated scaling of bulk energy gap and $c_{\rm{eff}} \approx \ln(2)$ for the QCPs at finite $\sigma_{w}$ when $\lambda<1$ and $\lambda>1$, respectively.} 
\label{fig_sigmaw}
\end{figure*}

\section{Estimation of probability of having a topological unit cell using normal probability distribution}

In this section, we estimate the probability of having a topological unit cell and show that this probability can also distinguish the topological phase from the trivial phase.

\begin{figure*}[t]
    \centering
\includegraphics[width=0.9\linewidth]{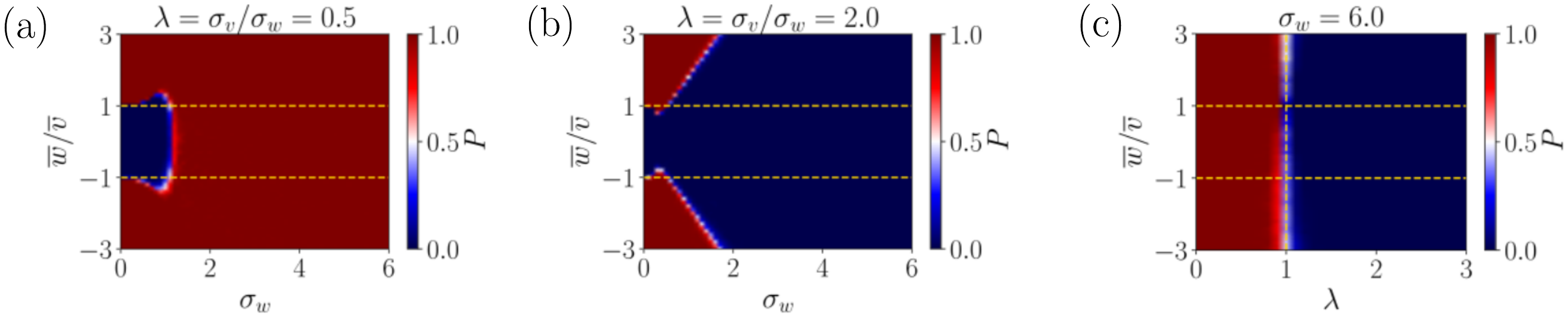}
\caption{{\textbf{Polarization for uniform probability distributions:}} (a,b) Polarization $P$ as a function of ${\overline{w}}/{\overline{v}}$ and $\sigma_{w}$ when $v$ and $w$ are chosen from uniform probability distributions where $v \in [{\overline{v}}-\sqrt{3} \sigma_{v}, {\overline{v}}+\sqrt{3} \sigma_{v}]$ and $w \in [{\overline{w}}-\sqrt{3} \sigma_{w}, {\overline{w}}+\sqrt{3} \sigma_{w}]$. In (a) $\lambda=\sigma_{v}/\sigma_{w}=0.5$ and in (b) $\lambda=2.0$. (c) Polarization $P$ as a function of ${\overline{w}}/{\overline{v}}$ and $\lambda$ when $v$ and $w$ are chosen from uniform probability distributions where $\sigma_{w}=6.0$. In all plots, $L=200$ and $P$ is averaged over $200$ realizations.
}
\label{fig_boxdist}
\end{figure*}

We consider the following normal probability distributions of intra-cell ($v$) and inter-cell ($w$) hopping parameters:
\begin{align}
p_{1}(v) &= \frac{1}{\sqrt{2 \pi \sigma_{v}^{2}}} \exp \left(-\frac{(v-{\overline{v}})^{2}}{2 \sigma_{v}^{2}} \right),\\
p_{2}(w) &= \frac{1}{\sqrt{2 \pi \sigma_{w}^{2}}} \exp \left(-\frac{(w-{\overline{w}})^{2}}{2 \sigma_{w}^{2}} \right).
\end{align}
Now, in the limit $\lambda \to 0$, $\sigma_{v} \to 0$, and thus a unit cell is topological if the inter-cell hopping parameter $w$ associated with that unit cell satisfies $|w|>|{\overline{v}}|$. Therefore, the probability of having a topological unit cell when $\lambda \to 0$ turns out to be
\begin{align}\label{eq_ptop_zero}
    p_{\rm{top}} &= \int_{-\infty}^{-|{\overline{v}}|} dw p_{2} (w) + \int_{|{\overline{v}}|}^{\infty} dw p_{2} (w) \nonumber \\
    &= 1- \int_{-|{\overline{v}}|}^{|{\overline{v}}|} dw p_{2} (w) \nonumber \\
    &= 1- \frac{1}{2} {\rm{erf}}\left( \frac{|{\overline{v}}|+ |{\overline{w}}|}{\sqrt{2} \sigma_{w}}\right) - \frac{1}{2} {\rm{erf}}\left( \frac{|{\overline{v}}|- |{\overline{w}}|}{\sqrt{2} \sigma_{w}}\right),
\end{align}
where the function ${\rm{erf}}(x)= \frac{2}{\sqrt{\pi}} \int_{0}^{x} dy \exp(-y^{2})$ is the error function. In contrast, when $\lambda \to \infty$, $\sigma_{w} \to 0$ and a unit cell is topological if the intra-cell hopping parameter $v$ associated with that unit cell satisfies $|v|<|{\overline{w}}|$. Thus, the probability of having a topological unit cell in the limit $\lambda \to \infty$ can be calculated as
\begin{align}\label{eq_ptop_infinity}
    p_{\rm{top}} &=  \int_{-|{\overline{w}}|}^{|{\overline{w}}|} dv p_{1} (v) \nonumber \\
    &= \frac{1}{2} {\rm{erf}}\left( \frac{|{\overline{w}}|+ |{\overline{v}}|}{\sqrt{2} \sigma_{v}}\right) + \frac{1}{2} {\rm{erf}}\left( \frac{|{\overline{w}}|- |{\overline{v}}|}{\sqrt{2} \sigma_{v}}\right).
\end{align}

In the clean limit $\sigma_{v} \to 0$, $\sigma_{w} \to 0$, $p_{\rm{top}}$ reduces to
\begin{equation}
p_{\rm{top}}= \begin{cases}
1 \text{~~~~~ for $|w|>|v|$ ~~(topological),}\\
0 \text{~~~~~ for $|w|<|v|$ ~~(trivial),}\\
1/2 \text{~~ for $|w|=|v|$~~~(QCP).}
\end{cases}
\end{equation}
Therefore, $p_{\rm{top}}=1/2$ corresponds to QCP separating the topological and trivial phases.

Further, for a finite $\lambda$, a unit cell is topological if the associated $w$ and $v$ satisfy $|w|>|v|$ and thus $p_{\rm{top}}$ can be obtained from the following equation:
\begin{align}\label{eq_ptop_detail}
p_{\rm{top}} &= \int_{-\infty}^{\infty} dw p_{2}(w) \int_{-|w|}^{|w|} dv p_{1}(v) \nonumber \\
&=\frac{1}{2} \int_{0}^{\infty} dw p_{2}(w) \left(  {\rm{erf}}\left( \frac{w+ {\overline{v}}}{\sqrt{2} \sigma_{v}}\right) + {\rm{erf}}\left( \frac{w- {\overline{v}}}{\sqrt{2} \sigma_{v}}\right) \right) \nonumber \\ 
& + \frac{1}{2}  \int_{-\infty}^{0} dw p_{2}(w) \left( {\rm{erf}}\left( \frac{-w+ {\overline{v}}}{\sqrt{2} \sigma_{v}}\right) + {\rm{erf}}\left( \frac{-w- {\overline{v}}}{\sqrt{2} \sigma_{v}}\right) \right).
\end{align}

In Fig.~\ref{fig_comparison}(a-c), we compare numerical and SDRG results with the analytical estimation of $p_{\rm{top}}$. Here, we observe that the numerically obtained polarization $P$ shows a transition from $P \sim 1$ to $P \sim 0$ at $\lambda=1$, while the quantity $|\overline{w_{\rm{eff}}}/\overline{v_{\rm{eff}}}|$ in SDRG method also undergoes a change from $|\overline{w_{\rm{eff}}}/\overline{v_{\rm{eff}}}|>1$ to $|\overline{w_{\rm{eff}}}/\overline{v_{\rm{eff}}}|<1$ at $\lambda=1$. Interestingly, analytical estimate of $p_{\rm{top}}$ captures the QCP at $\lambda=1$ where we find $p_{\rm{top}} \sim 1/2$. Thus, the analytical estimation of $p_{\rm{top}}$ is consistent with the numerical and SDRG results.

\section{$\sigma_{w}$-driven topological phase transitions}

While we have explicitly discussed $\lambda$-driven topological phase transitions in main text, topological phase transitions can also be driven by the parameter $\sigma_{w}$. To study $\sigma_{w}$-driven transitions, we explore the variation of the configuration averaged polarization $P$ with $\sigma_{w}$ and ${\overline{w}}/{\overline{v}}$ when $\lambda<1$ and $\lambda>1$, as discussed below.

When $\lambda<1$ and ${|\overline{w}}|<{|\overline{v}|}$, polarization shows a transition from trivial phase ($P \sim 0$) to topological phase ($P \sim 1$) as we tune $\sigma_{w}$ (see Fig.~\ref{fig_sigmaw}(a)). The QCP at finite $\sigma_{w}$, where the topological phase transition occurs, shows activated scaling of energy-gap $\epsilon \sim \exp(-a \sqrt{L})$ (where $a={\rm{constant}}$), as evident from the collapse of probability distribution $\tilde{p}$ of $\epsilon$ in Fig.~\ref{fig_sigmaw}(b). Also, the central charge associated with this QCP is $c_{\rm{eff}}=\ln(2)$ which is obtained from the logarithmc scaling of entanglement entropy $S_{l}$ (see Fig.~\ref{fig_sigmaw}(c)). These observations confirm that $\sigma_{w}$-driven topological phase transition for $\lambda<1$ and ${|\overline{w}}|<{|\overline{v}|}$ occurs at an infinite randomness fixed point (IRFP).

Similarly, When $\lambda>1$ and ${|\overline{w}}|>{|\overline{v}|}$, there is another QCP at a finite $\sigma_{w}$ where polarization exhibits a topological ($P \sim 1$) to trivial ($P \sim 0$) transition (see Fig.~\ref{fig_sigmaw}(d)). This QCP is also an IRFP, which is confirmed by activated scaling of $\epsilon$ and effective central charge $c_{\rm{eff}}=\ln(2)$ (see Fig.~\ref{fig_sigmaw}(e,f)).

\section{Strong disorder-driven topological phase transitions with uniform probability distributions}
In this section, we discuss strong disorder-driven topological phase transitions considering the uniform probability distributions. Here, the intra-cell ($v$) and inter-cell ($w$) hopping parameters are chosen from the following uniform probability distributions:
\begin{subequations}
\begin{equation}
    p_{1}(v) =\begin{cases}
        \frac{1}{2 \sqrt{3} \sigma_{v}},&\text{~~~~for $({\overline{v}}-\sqrt{3} \sigma_{v}) \leq v \leq ({\overline{v}}+\sqrt{3} \sigma_{v})$},\\
        ~~~0 ,&\text{~~~~otherwise,}
    \end{cases}
\end{equation}
\begin{equation}
    p_{2}(w) =\begin{cases}
        \frac{1}{2 \sqrt{3} \sigma_{w}},&\text{~~~~for $({\overline{w}}-\sqrt{3} \sigma_{w}) \leq w \leq ({\overline{w}}+\sqrt{3} \sigma_{w})$},\\
        ~~~0 ,&\text{~~~~otherwise,}
    \end{cases}
\end{equation}
\end{subequations}
where ${\overline{v}}$, ${\overline{w}}$ are the means of the two distributions and the standard deviations are $\sigma_{v}$, $\sigma_{w}$ respectively. Similar to the situation with normal distribution, here also the chain with strong disorder (large $\sigma_{w}$) becomes topological ($P \sim 1$) when $\lambda=\sigma_{v}/{\sigma_{w}}<1$ (see Fig.~\ref{fig_boxdist}(a)), while it becomes trivial ($P \sim 0$) when $\lambda>1$ (see Fig.~\ref{fig_boxdist}(b)). Thus, for large $\sigma_{w}$, topological to trivial phase transition occurs at $\lambda=1$, irrespective of the value of ${\overline{w}}/{\overline{v}}$ (see Fig.~\ref{fig_boxdist}(c)).

%\end{document}

\end{document}